\documentclass[letterpaper, 10 pt, conference]{ieeeconf}  
\usepackage{comment}       
\usepackage{cite}
\usepackage{amsmath, amssymb}
\usepackage{graphicx}
\usepackage{ragged2e}
\usepackage{threeparttable}
\usepackage{placeins}
\usepackage{dblfloatfix}
\usepackage{url}
\usepackage[hidelinks]{hyperref}

\IEEEoverridecommandlockouts                              

\title{\LARGE \bf
nASR: An End-to-End Trainable Neural Layer for Channel-Level EEG Artifact Subspace Reconstruction in Real-Time BCI
}

\author{Shantanu Sarkar$^{1}$ and Jose L. Contreras-Vidal$^{2}$
\thanks{*This work was supported by NSF IUCRC BRAIN award \# 2137255}
\thanks{$^{1}$Shantanu Sarkar is a doctoral candidate, Dept. of ECE, Univ. of Houston, Houston, TX, USA, {\tt\small shantanu75@gmail.com}}%
\thanks{$^{2}$Jose L. Contreras-Vidal is with Faculty of ECE, Univ. of Houston, Houston, TX, USA, {\tt\small jlcontreras-vidal@uh.edu}}%
}

\begin{document}

\maketitle
\thispagestyle{empty}
\pagestyle{empty}

\begin{abstract}
Electroencephalogram (EEG) signals are highly susceptible to artifacts, resulting in a low signal-to-noise ratio, which makes extraction of  meaningful neural information challenging. Artifact Subspace Reconstruction (ASR) is one of the most widely used artifact filtering techniques in EEG-based BCI applications, owing to its real-time applicability. ASR reconstructs artifact-free signals by operating in Principal Component (PC) space within sliding windows. However, ASR performance is critically sensitive to its threshold parameter -- an incorrect threshold risks removing task-relevant neural features alongside artifacts. Furthermore, since PCs are linear combinations of all channels, subspace reconstruction in PC space may alter the underlying data structure, potentially discarding essential neural information.
\\ 
To address these limitations, we propose nASR, a novel end-to-end trainable Keras layer that jointly optimizes artifact rejection and downstream decoding. nASR introduces two trainable threshold parameters: K, which governs artifact detection in PC variance space, and L, which quantifies eigen-spread to pinpoint the primary artifact--contributing channels, enabling selective channel-level reconstruction that preserves clean channel information.
\\
An ablation study comprising five model variants (m01--m05), evaluated across human subject data from the BCI Competition IV Dataset 1, confirms that nASR variants consistently outperform traditional ASR on test classification metrics, while achieving a $>20\times$ reduction in inference time, making nASR a strong candidate for real-time BCI applications demanding both low latency and high decoding performance.
\end{abstract}
\section{INTRODUCTION}
Electroencephalography (EEG) is a non-invasive technique used to record brain activity via scalp electrodes, providing spatial, temporal, and spectral patterns related to sensory and cognitive processes\cite{Cohen2017}. Its non-invasive nature, safety, affordability, and real-time measurement of neuronal activity make it widely used in clinical and neuroscience research, and popular for brain-computer interface (BCI) applications\cite{Sugden2023, Craik2019}. However, EEG recordings are influenced by physiological artifacts (eye movements, muscle activity) and non-physiological noises (electromagnetic and powerline interference), reducing the signal-to-noise ratio (SNR) and limiting reliable analyses \cite{Jiang2019, Rashmi2022, Kilicarslan2021}. Effective denoising is therefore essential, and remains challenging when artifact frequencies overlap with meaningful EEG activity \cite{Kilicarslan2021}.
\\
EEG denoising approaches fall into three main categories: frequency-based filtering, adaptive filtering, and blind source separation (BSS). In frequency-based filtering, specific frequency bands are suppressed using tools such as band-pass (BPF), low-pass (LPF), high-pass (HPF), notch, or wavelet-based filters; however, it cannot distinguish artifacts from features sharing the same band. Adaptive filtering methods, such as H-Infinity filters, have shown strong potential for real-time artifact removal but require a reference noise signal that is not always available \cite{Kilicarslan2016, Kilicarslan2019}. Blind source separation (BSS) methods, such as Independent Component Analysis (ICA) and Principal Component Analysis (PCA), decompose multichannel EEG into components under a linear mixing model \cite{Jiang2019,Rashmi2022}; ICA is widely used but is computationally costly, less effective on transient artifacts, and often requires manual component selection \cite{Jiang2019,Rashmi2022,Chang2020}.
\\
The Artifact Subspace Reconstruction (ASR) method identifies and suppresses artifacts by projecting EEG data within sliding windows into a principal component (PC) space derived from the covariance of clean reference data, removing components that exhibit abnormally high variance \cite{Chang2020}. ASR suits real-time use due to its low overhead and minimal manual intervention. However, its performance is highly sensitive to the thresholding parameter: an incorrect threshold can strip meaningful EEG features along with artifacts \cite{Chang2020, Anders2020, Wojcik2023}. Additionally, because PCs mix all channels, reconstruction alters the data structure \cite{Ibraheem2014}, potentially distorting task-relevant features.
\\
No existing method jointly optimizes artifact rejection thresholds and downstream decoding within a single trainable framework. We propose nASR, a neural-network layer that reimagines EEG artifact subspace reconstruction as a learnable process. nASR introduces trainable thresholding in the PC space, followed by eigen-spread thresholding \cite{Dapena2012} to detect and reconstruct artifact-contaminated channel subspaces, leveraging volume conduction \cite{Cohen2017}. nASR learns to pinpoint and reconstruct only the primary artifact-contributing channels,  preserving clean channels. By embedding artifact rejection directly into the network, it optimizes the detection threshold and decoder jointly, eliminating the separate hand-tuned preprocessing stage.
\\
The key contributions of this paper are: (1) a novel end-to-end trainable layer for channel-level EEG artifact subspace reconstruction with learnable dual thresholds; (2) channel-level artifact identification and selective reconstruction preserving clean channel integrity; (3) a comprehensive ablation study validating each design component; and (4) a Pareto analysis demonstrating inference time reduction over traditional ASR.
\section{MATERIALS AND METHODS}
\subsection{Dataset}
\begin{figure}[!t]
    \centering
    \includegraphics[width=\linewidth]{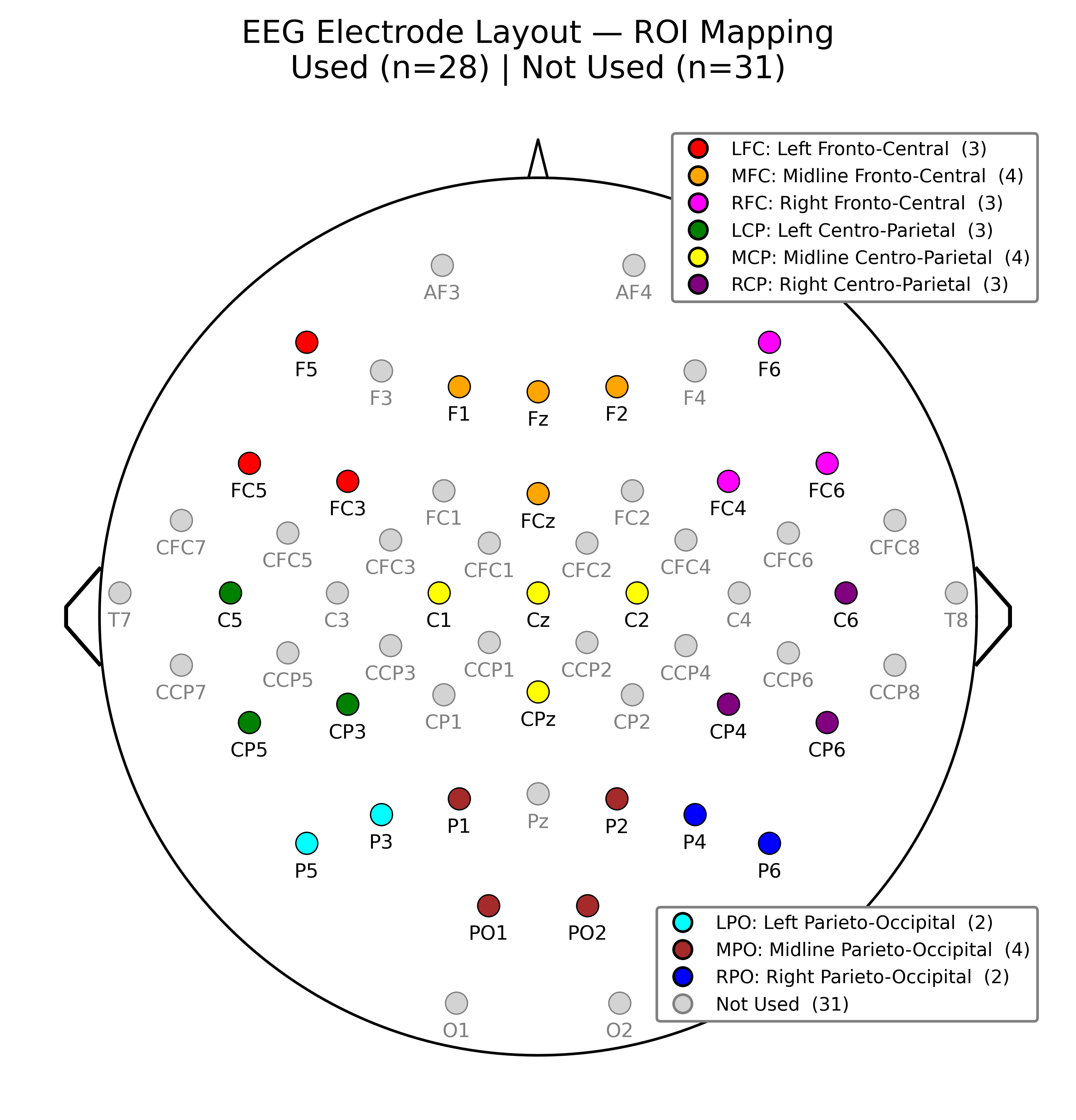}
    \caption{EEG electrode layout illustrating the 28 selected channels grouped into nine ROIs. Unused channels ($n = 31$) are shown in gray.}
    \label{Fig1}
\end{figure}

Our goal is to reconstruct contaminated EEG subspaces in real-time to improve motor imagery decoding. Hence, to develop the algorithm, we used the BCI Competition IV (Dataset 1) \cite{Blankertz2007}. The dataset contains 59-channel EEG recordings from seven subjects performing binary MI tasks (`a', `b', `g', and `f' are from healthy subjects, and the remaining `c', `d', and `e' are synthetically generated data). For each subject, binary classes of motor imagery were selected from one of three different options: left hand, right hand, and foot (selected by the subject). The dataset is available in two different sets -- labeled (calibration) and unlabeled (evaluation), in two forms: the original 1000 Hz recordings, which were band-pass filtered between 0.05 and 200 Hz, and a down-sampled 100 Hz version derived through low-pass filtering and block averaging. 
\\
For our study, we used the 100 Hz sampled dataset from healthy subjects (`a', `b', `g', and `f'). Among the 59 available EEG channels, we used 28 channels from 9 regions of interest (ROI) as shown in Fig.~\ref{Fig1}. We mainly excluded ROIs from the outer edge, such as Anterior Frontal, Occipital, and Left/Right Temporal to limit EMG contamination, as EMG contamination increases with distance from the vertex \cite{Pope2022}.
\subsection{Preprocessing}
We further preprocessed the channel-wise EEG data using a zero-phase-shift 6th-order Butterworth BPF with low and high cutoffs of 0.5 Hz and 30 Hz, respectively. We selected the high cutoff frequency of 30 Hz to suppress the gamma band, thereby reducing high-frequency noise and, more specifically, minimizing EMG contamination. EMG artifacts are known to progressively dominate EEG signals at frequencies above $\sim25$ Hz \cite{Pope2022}. This choice was made to ensure that the decoding process reflects neural activity rather than inadvertent EMG contributions arising during actual hand or leg movements.
\\
From the continuous EEG recordings, we generated sliding windows of 256 samples (2.56 s) with a step size of 20 samples (200 ms). The 256-sample window length was chosen because it provides a minimum resolvable frequency of approximately 0.39 Hz, ensuring that all frequency components above $\sim0.5$ Hz are preserved in the spectral representation.
\\
For the baseline, we applied traditional Artifact Subspace Reconstruction (ASR) using the asrpy Python package (version 0.0.8). The ASR was initialized with the dataset’s sampling rate and a standard cutoff threshold of 20. We calibrated the ASR model using 60 seconds of resting-state data (20 seconds repeated three times), consistent with the recommended minimum calibration duration of approximately 1 minute of clean resting EEG data \cite{Mullen2015}. After calibration, the fitted ASR model was used to transform all sliding windows of the EEG recording, producing the baseline cleaned signal used for comparisons. 
\subsection{Neural Artifact Subspace Reconstruction (nASR)}
Here, we propose a neural-network–based ASR (nASR) layer, with the novel capability of identifying artifact-contributing channels, implemented as a trainable Keras layer that operates directly within the computational graph, with thresholding parameters learned as weights. These trainable thresholding parameters enable optimal thresholding, ensuring that important neural features are not removed. Additionally, in nASR, we reconstruct the contaminated channel rather than the subspace in PC space, thereby retaining the features in good channels intact.  
\\
Given an EEG input shaped (B, C, W), where B is the batch size, C is the total number of channels, and W is the number of samples in the sliding window, the input is first channel-wise Z-score normalized using clean reference statistics ($\mu$ and $\sigma$ per channel), such that each channel has zero mean and unit variance. The nASR layer then computes a mean-centered covariance matrix over the non-overlapping portion of each sliding window, producing a (B, C, C) estimate of the multichannel signal's spatial structure. By performing an eigen-decomposition, the model identifies how variance is distributed across the data's directions. A learnable threshold, bounded below by a fixed offset $K_{Offset}$, is applied to the per-channel eigenvalue projections to identify channels exhibiting artifact-like activity. These binary artifact decisions are implemented via a Straight-Through Estimator (STE), which produces hard binary masks for the forward pass, while propagating gradients through the threshold during training, enabling end-to-end optimization.
\\
To identify the clean reference segments, we first computed the minimum and maximum thresholds as $\mu \pm3.5\sigma$ across all sliding windows. A window is identified as clean only if the EEG signals of all channels remained within this bounded range; windows violating this criterion in even a single channel were discarded. From the retained clean windows, channel-wise reference statistics ($\mu$ and $\sigma$ per channel) were then computed, serving as the baseline for subsequent artifact reconstruction.
\\
In contrast to traditional ASR, we introduced an additional thresholding step to identify channels exhibiting excessive noise spread. Channels flagged as noisy via a second trainable threshold parameter are reconstructed solely using a weighted average of their spatially proximate clean neighbors. If all neighboring channels are also identified as noisy, the reconstruction falls back to a weighted mean computed from the remaining artifact-free channels. Each component of the proposed algorithm is described in Section~\ref{sec:nasr}, where the full nASR architecture is presented in detail.
\subsection{Ablation Study - Downstream MI Classification}
To assess the contribution of each component within the nASR framework, an ablation study was conducted using motor imagery classification with EEGNet \cite{EEGNet2018} as the downstream decoder. Five experimental conditions were evaluated: a baseline pipeline (m01: conventional ASR followed by average re-referencing) and four ablation variants (m02–m05), each systematically toggling a specific design choice within the nASR framework. Three key design elements were examined: (1) artifact reconstruction using spatially proximate clean neighboring channels versus the weighted mean of all artifact-free channels, (2) covariance estimation over the full sliding window versus only the non-overlapping segment, and (3) fixed versus learnable weighted reconstruction of artifact-contaminated channels. Table~\ref{Table1} summarizes the configuration settings for each experimental condition. This design isolates the contribution of each module and reveals how individual components interact within the complete nASR pipeline.
\begin{table}[ht]
\centering
\begin{threeparttable}
\caption{Ablation study configuration}
\label{Table1}
\renewcommand{\arraystretch}{1.5}
\begin{tabular}{c c c c c}
\hline
\textbf{Model}    & \textbf{ASR vs.}     & \textbf{Reconstruct} & \textbf{Covariance}  & \textbf{Reconstruct} 
\\
\textbf{Variants} & \textbf{nASR} & \textbf{(Neighbors)} & \textbf{(Full Win.)} & \textbf{(Weighted)} \\

\hline
m01 & ASR  & --    & --     & --    \\
m02 & nASR & False & False  & True  \\
m03 & nASR & True  & False  & True  \\
m04 & nASR & True  & True   & True  \\
m05 & nASR & True  & False  & False \\
\hline
\end{tabular}
\begin{tablenotes}
\footnotesize
\item Note: Five experimental conditions -- m01 serves as the traditional ASR baseline; m02--m05 are nASR variants with specific components toggled to isolate their individual contributions.
\end{tablenotes}
\end{threeparttable}
\end{table}
\renewcommand{\arraystretch}{1.0}
\subsection{Training \& Validation Strategy}
For model training, we used the Adam optimizer (initial lr $= 10^{-3}$, clipnorm $= 1.0$, batch size $= 64$) for up to 250 epochs, with a 5-epoch linear warmup, learning-rate halving on validation-loss plateau (patience $= 5$, min $= 10^{-7}$), and early stopping (patience $= 50$) restoring the best weights. A 20\% dropout rate was applied throughout the EEGNet decoder during training. A combined loss function was employed, using Binary Cross-Entropy (BCE) and Dice Loss. The Dice Loss complements BCE by directly optimizing the overlap between predicted and true class distributions, improving robustness under class imbalance. Both loss components are integrated via an equal-weighted sum:
$$
L_{combined} =0.5\cdot L_{BCE} +0.5\cdot L_{Dice}. \eqno{(1)}
$$
The dataset was partitioned sequentially into training (60\%), validation (20\%), and test (20\%) subsets based on the temporal order of the sliding windows, ensuring minimum data leakage across splits.
\subsection{Experimental Platform}
All experiments were conducted on a workstation running Ubuntu Linux (kernel 6.5.0, x86\_64), equipped with a 32-core (64 logical threads) CPU clocked at 3.0 GHz, 270 GB of system RAM, and dual NVIDIA RTX 6000 Ada Generation GPUs, each with 48 GB of VRAM (96 GB total GPU memory), under driver version 545.23.08. Model training and inference were performed using Python 3.11.15, TensorFlow 2.16.1, and Keras 3.14.0.
\begin{figure}[!t]
    \centering
    \includegraphics[width=\linewidth]{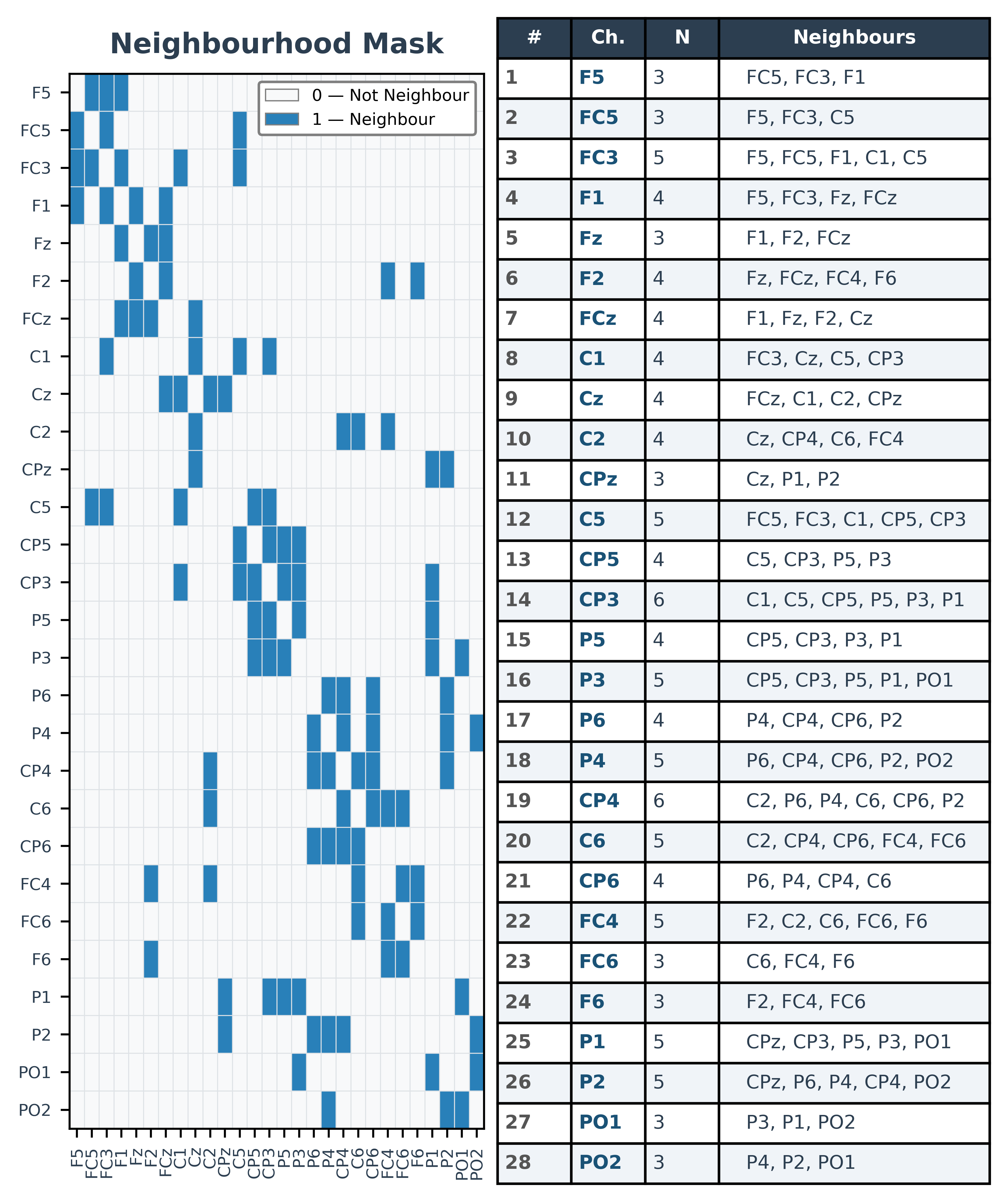}
    \caption{Neighborhood mask for the 28-channel EEG configuration. (Left) Binary heatmap where blue cells indicate neighboring channel pairs (L2 distance $<$ 0.05) and white cells indicate non-neighbors. (Right) Channel-wise neighbor list showing each channel index, label, number of neighbors (N), and corresponding neighbor channel labels in the 10–20 system.}
    \label{Fig2}
    \vspace{-10pt} 
\end{figure}
\section{$nASR$: Architecture and Algorithm}\label{sec:nasr}
\subsection{Covariance Estimation and Eigen-Decomposition}
Given sliding windows of EEG input $X \in \mathbb{R}^{B \times C \times W}$, where B, C, and W denote batch size, channels, and window length, respectively, the nASR layer first computes a mean-centered sliding window covariance matrix over the non-overlapping S samples of the sliding window. For each sliding window in the batch, the covariance matrix is computed as:
$$
    \text{Cov}_S = \frac{\mathbf{X}_S \cdot \mathbf{X}_S^T}{S - 1}, \eqno{(2)}
$$
where
\begin{itemize}
    \item $X_S \in \mathbb{R}^{C \times S}$: Mean-centered EEG of a non-overlapping segment.
    \item $\mathrm{Cov}_S \in \mathbb{R}^{C \times C}$: Channel-by-channel covariance matrix.
\end{itemize}

\noindent Eigen-decomposition of $\text{Cov}_S$ yields eigenvalues D and eigenvectors V, which together characterize the directional variance structure of the multichannel signal.
$$
\mathrm{Cov}_S = V \cdot D \cdot V^{T}, \eqno{(3)}
$$
where
\begin{itemize}
    \item $D \in \mathbb{R}^{C \times C}$: Diagonal matrix of eigenvalues.
    \item $V \in \mathbb{R}^{C \times C}$: Orthogonal matrix of eigenvectors.
\end{itemize}
\subsection{Artifact Threshold and Component Rejection}
Since the EEG input is channel-wise Z-score normalized using clean reference statistics, all channels are transformed to zero mean and unit variance ($\mu$ = 0, $\sigma^2$ = 1) in the normalized space. Consequently, the per-channel artifact threshold simplifies to a single learnable scalar, eliminating dimensional inconsistency between the mean and variance terms. The threshold is computed as:
$$
\mathrm{Th} = K + K_{\mathrm{Offset}} \quad (\text{under } \mu = 0,\ \sigma^2 = 1),\eqno{(4)}
$$
where 
\begin{itemize}
    \item $K_{\mathrm{Offset}} \geq 0.1$: Fixed lower bound (default $0.1$).
    \item $K \geq 0$: Learnable scalar initialized at $K_{\mathrm{init}} \approx 0.71$.
\end{itemize}

\noindent The learnable parameter K is initialized as $K_{\mathrm{init}} =(0.9)^2 - K_{\mathrm{Offset}}$, corresponding to a threshold at 90\% of the normalized standard deviation ($\sigma$ = 1), providing a conservative initial sensitivity to artifact-like variance prior to training.
\\
Eigenvector projections onto the threshold are computed as:
$$
\mathrm{Check}_c = \sum_{j} \left| \mathrm{Th} \cdot V_{(c,j)} \right|. \eqno{(5)}
$$
Components whose normalized eigenvalue difference exceeds the learned threshold are flagged as artifact contaminated. To keep the operation differentiable, a Straight Through Estimator (STE) is applied. The STE enables hard binary decisions during the forward pass while allowing gradients to flow through a smooth sigmoid approximation during backpropagation. The discard mask is computed as:
$$
\mathrm{Diff} = D - \mathrm{Check}. \eqno{(6)}
$$
$$
\mathrm{NormDiff} = \frac{\mathrm{Diff}}{\max\left( \mathrm{mean}(|\mathrm{Diff}|), \, \epsilon \right)}.\eqno{(7)}
$$
$$
\mathrm{Discard}_{\mathrm{s}} = \mathrm{Sigmoid}\left( \tau_D \cdot \mathrm{NormDiff} \right). \eqno{(8)}
$$
$$
\mathrm{Discard}_{\mathrm{h}} = \mathrm{round}\left( \mathrm{Discard}_{\mathrm{s}} \right). \eqno{(9)}
$$
$$
\mathrm{Discard} = \mathrm{Discard}_{\mathrm{s}} 
+ \mathrm{sg}\left( \mathrm{Discard}_{\mathrm{h}} - \mathrm{Discard}_{\mathrm{s}} \right), 
\eqno{(10)}
$$
where
\begin{itemize}
    \item $\mathrm{Diff} \in \mathbb{R}^{B \times C}$
    \item $\mathrm{NormDiff} \in \mathbb{R}^{B \times C}$
    \item $\mathrm{Discard}_{s} \in [0,1]$
    \item $\mathrm{Discard}_{h} \in \{0,1\}$
    \item $\mathrm{sg}(\cdot)$ denotes the stop-gradient operator
    \item $\tau_D = 20$
    \item $\epsilon = 10^{-6}$
\end{itemize}
\subsection{Channel-Level Noise Classification} 
Channel-Level Noise Classification is done via an additional learnable threshold L. Each channel's artifact spread is defined as the sum of squared projections onto discarded eigenvectors:
$$
\mathrm{Spread}_c = \sum_{j} \left( \mathrm{Discard}_j \cdot V_{(c,j)} \right)^2. 
\eqno{(11)}
$$
$$
\mathrm{Noise}_{s,c} = \mathrm{Sigmoid}\left( \tau_L \left( \mathrm{Spread}_c - L \right) \right). 
\eqno{(12)}
$$
$$
\mathrm{Noise}_h = \mathrm{round}\left( \mathrm{Noise}_s \right). 
\eqno{(13)}
$$
$$
\mathrm{Noise} = \mathrm{Noise}_s + \mathrm{sg}\left( \mathrm{Noise}_h - \mathrm{Noise}_s \right), 
\eqno{(14)}
$$
where 
\begin{itemize}
    \item $\mathrm{Spread}_c \in [0,1]$
    \item $\mathrm{Noise}_s \in [0,1]$
    \item $\mathrm{Noise}_h \in \{0,1\}$
    \item $\mathrm{sg}(\cdot)$ denotes the stop-gradient operator
    \item $\tau_L = 20$
\end{itemize}
\begin{figure*}[t]
    \centering
    \includegraphics[width=\textwidth]{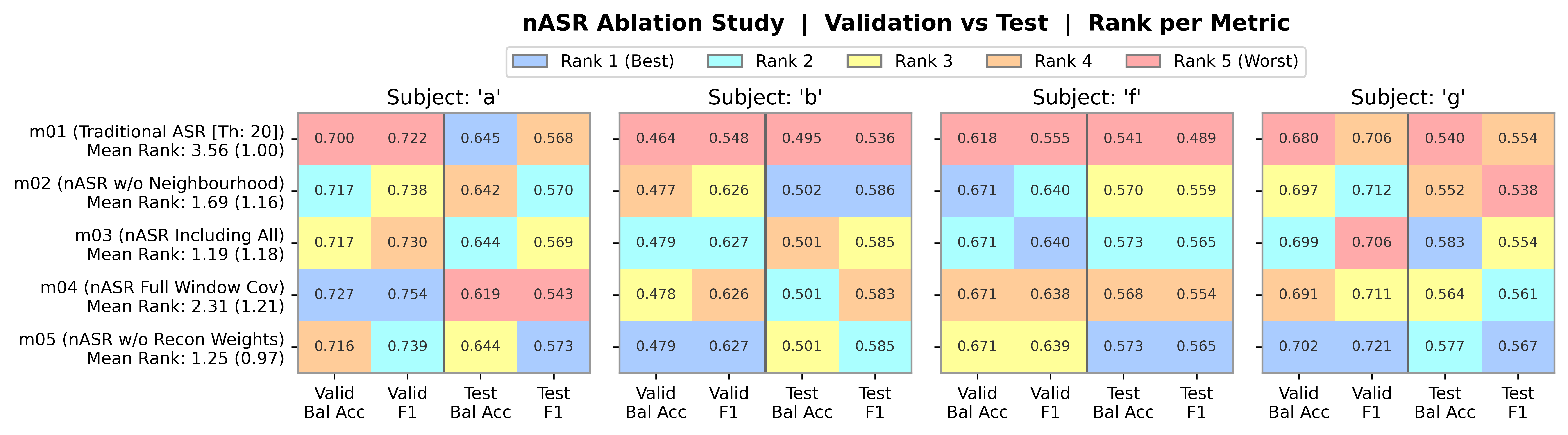}
    \vspace{-5pt}
    \caption{Rank-based performance heatmap across validation and test splits
    for four subjects (`a', `b', `f', and `g'). Cell color indicates
    within-column rank, from Rank~1 (blue, best) to Rank~5 (red, worst).
    Metrics are balanced accuracy and F1-score; each model's mean rank (SD)
    is shown beside its label.}
    \label{Fig3}
    \vspace{-5pt}
\end{figure*}
\subsection{Subspace Reconstruction} 
As channels with $\mathrm{Noise}_c \approx 1$ are classified as artifact-contaminated, we defined the clean-channel mask as:
$$
\mathrm{Good}_c = 1 - \mathrm{Noise}_c. \eqno{(15)}
$$
The safe number of clean channels to compute average is computed as:
$$
\mathrm{nGood}_{\mathrm{safe}} = \max \left( \sum_{c} \mathrm{Good}_c ,\, 1 \right). \eqno{(16)}
$$
The noisy channels are replaced using the average of the clean channels as:
$$
\mathrm{CleanAv} = \frac{\sum_{c} \left( X_c \cdot \mathrm{Good}_c \right)}{\mathrm{nGood}_{\mathrm{safe}}}. 
\eqno{(17)}
$$
$$
\tilde{X}_c = X_c \cdot \mathrm{Good}_c + \mathrm{CleanAv} \cdot \mathrm{Noise}_c. 
\eqno{(18)}
$$
For the neighborhood-based reconstruction, we used a precomputed spatial neighborhood adjacency matrix $A \in \mathbb{R}^{C \times C}$, derived from electrode proximity \cite{Sarkar2025}. This neighborhood adjacency matrix maps each EEG channel to its nearest neighbors (L2 distance $<$ 0.05) in the 10–20 system \cite{Bocker1994}. The neighborhood mask ($A$) used in this study is illustrated in Fig.~\ref{Fig2} as a heatmap, along with a table showing each channel and its corresponding neighboring channels. If neighborhood-based reconstruction is enabled, with adjacency matrix $A$, the reconstruction is refined as:
$$
\mathrm{Recon}_c 
= \mathrm{Good}_c \cdot \tilde{X}_c 
+ \mathrm{Noise}_c \cdot 
\frac{\sum_{k} \left( A_{(c,k)} \cdot \tilde{X}_k \right)}
{\max\left( \sum_{k} A_{(c,k)}, \, 1 \right)}.
\eqno{(19)}
$$
The two learnable parameters, $K$ and $L$, are jointly optimized with the classification objective, allowing artifact rejection thresholds to adapt for optimal model performance. Along with the filtered signal, the Boolean noise mask: $\mathrm{Noise}_h \in \{0,1\}^{B \times C \times 1}$  is also returned for optional downstream use.
\subsection{Additional Preprocessing Layers} 
Post-nASR, we added two custom layers: the Weighted Reconstruction Layer and the Average Re-Referencing Layer, which are discussed below.
\subsubsection{Weighted Reconstruction Layer} 
To refine the reconstructed EEG signals, a channel-wise learnable scaling weight is applied exclusively to the artifact-contaminated channels identified by the noise mask. Let $\hat{X} \in \mathbb{R}^{C \times W}$ denote the reconstructed EEG for a single window, and $M \in \{0,1\}^{C \times 1}$ the binary noise mask identifying artifact-contaminated channels. A trainable non-negative scaling weight vector $\mathrm{ScalingW} \in \mathbb{R}_{+}^{C}$ is introduced to scale the contributions of masked (reconstructed) channels. The mask-gated scaling weight is computed as:
$$
\mathrm{ScalingW}_M = M \odot \mathrm{ScalingW}. \eqno{(20)}
$$
The final reconstructed signal combines the original clean channels (unmasked) with the weighted reconstructed channels (masked):
$$
X_{\mathrm{final}} 
= \hat{X} \odot (1 - M) 
+ \hat{X} \odot \mathrm{ScalingW}_M. 
\eqno{(21)}
$$
\subsubsection{Average Re-Referencing Layer} 
To improve signal consistency across channels, an average re-referencing operation is applied — the most widely adopted referencing technique in EEG analysis. The advantage of average reference is that, over a closed spherical surface, the outward positive and negative currents cancel, yielding a net potential near zero \cite{Yao2019}. The reference signal is computed as the mean across all C channels at each time point:
$$
r_t = \frac{1}{C+1} \sum_{c=1}^{C} x_{(c,t)}. \eqno{(22)}
$$
The re-referenced signal is then obtained by subtracting the common reference from each channel:
$$
\tilde{x}_{(c,t)} = x_{(c,t)} - r_t, \eqno{(23)}
$$
where $x_{(c,t)}$ denotes the EEG potential of channel $c$ at time $t$, and $\tilde{x}_{(c,t)}$ is the corresponding re-referenced signal.
\section{RESULTS}

\subsection{Ablation Study: Classification Performance}
Fig.~\ref{Fig3} reports the validation and test performance (Balanced Accuracy and F1-Score) of all five models across four subjects (`a', `b', `f', and `g'), with each model's mean rank aggregated across subjects and splits. The baseline m01 (Traditional ASR, Th:20) was weakest overall (mean rank 3.56), ranking last on most test metrics. Although often competitive on validation (e.g., 0.700 on `a'), its test performance dropped sharply (0.645), indicating overfitting.
\\
All learnable nASR variants outperformed the baseline. m03 (Including All) and m05 (without Reconstruction Weights) were best and most consistent (mean ranks 1.19 and 1.25; SD $\leq$ 1.18), followed by m02 (1.69). m04 (Full Window Covariance) ranked mid-pack (2.31): despite the best validation on `a' (0.727), it had the worst test performance there (0.619), mirroring the baseline's overfitting and indicating that full-window covariance harms generalization. Overall, all nASR variants outperformed the baseline (mean rank $\leq$ 2.31 vs.\ 3.56), demonstrating that the learnable nASR formulation generalizes more effectively than traditional ASR, provided the reconstruction is appropriately constrained.
\begin{figure}[!t]
    \centering
    \vspace{-5pt}  
    \includegraphics[width=\linewidth]{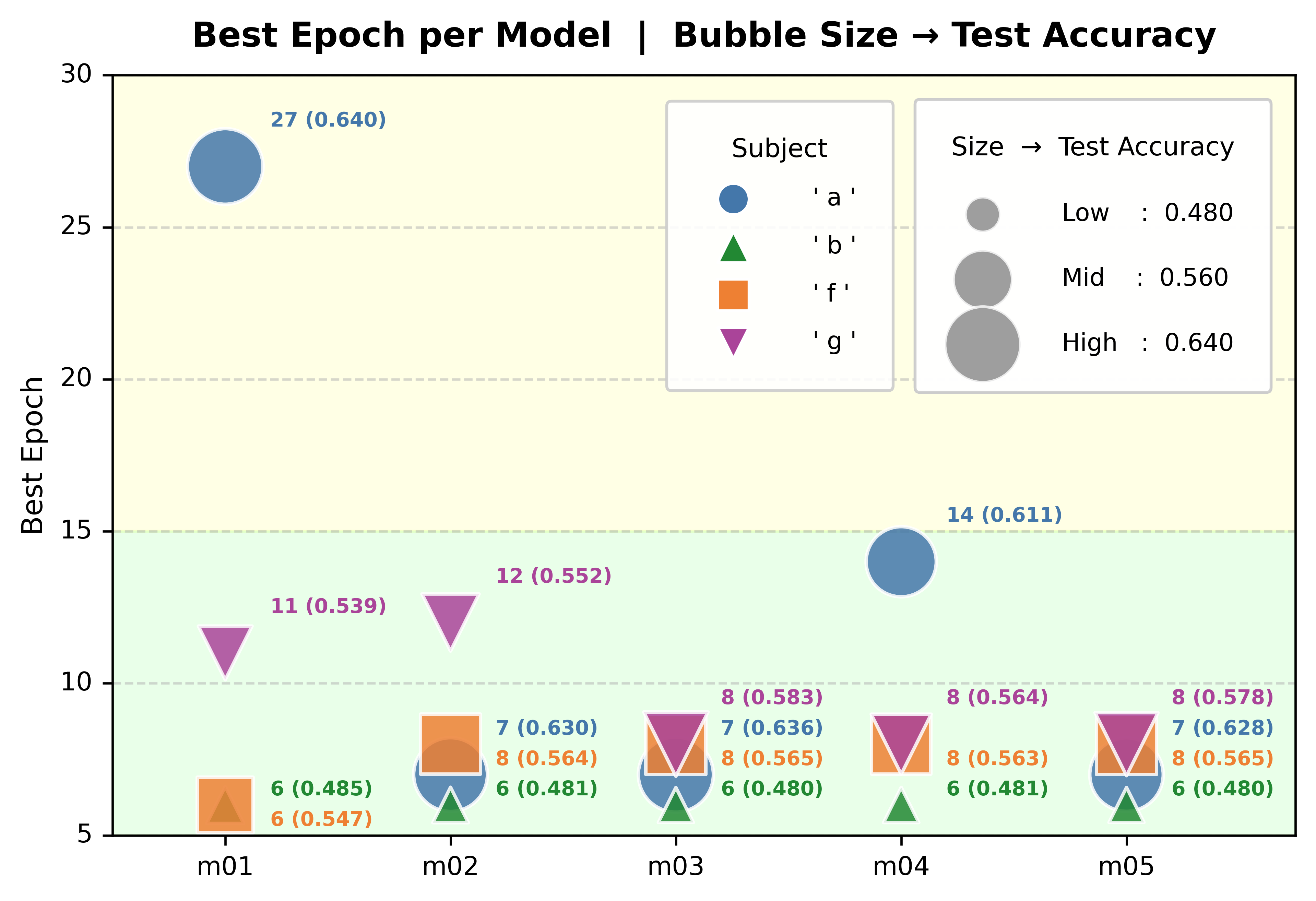}
    \caption{Best training epoch per model for four subjects: `a' (circle), `b' (triangle up), `f' (square), and `g' (triangle down). Bubble size encodes test accuracy. Background shading indicates convergence zones: Good (5--15) and Late ($>$15 epochs). Annotated values show best epoch (test accuracy).}
    \label{Fig4}
\end{figure}
\begin{figure}[!ht]
    \centering
    \vspace{-5pt}  
    \includegraphics[width=\linewidth]{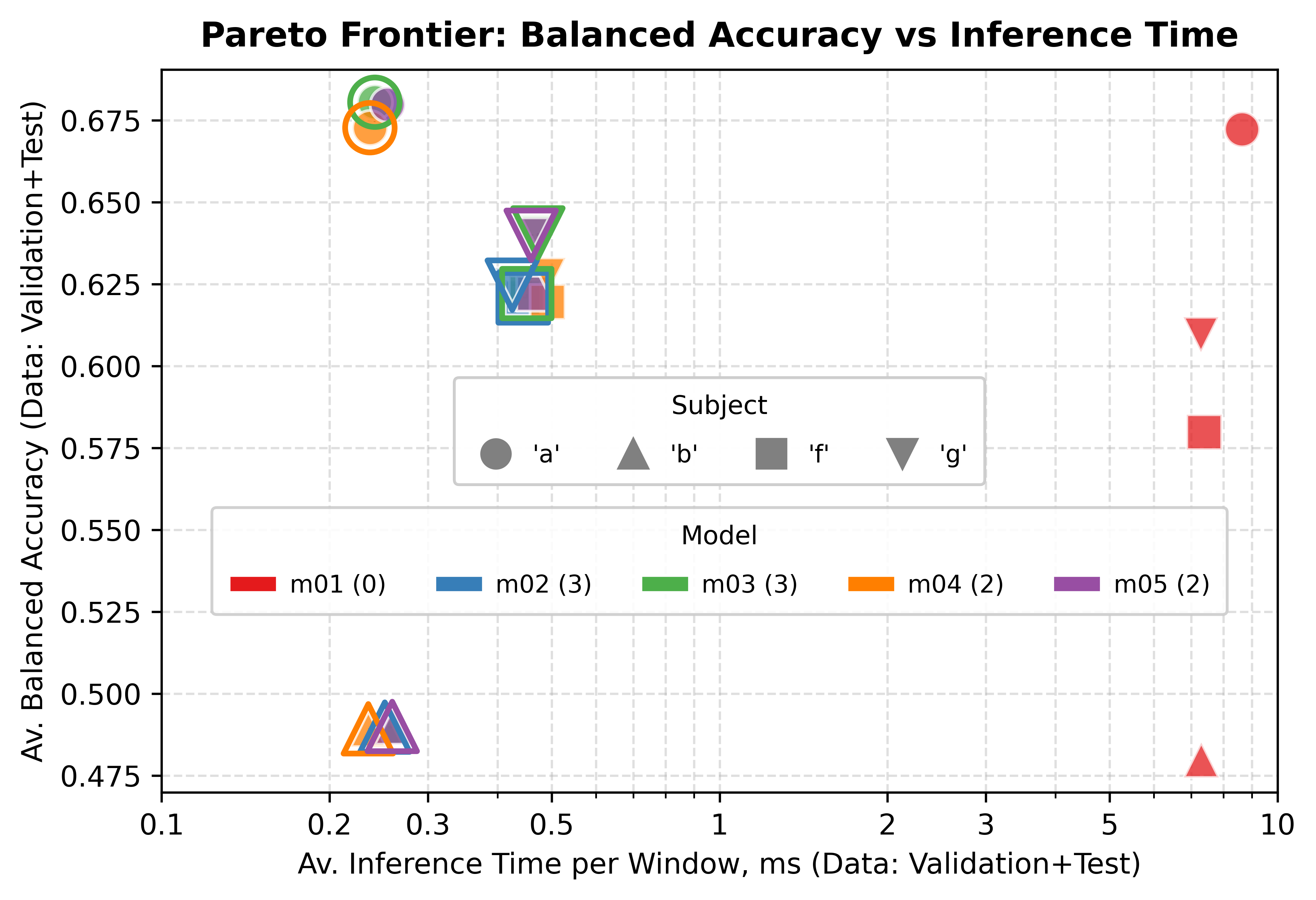} 
    \caption{Pareto frontier of balanced accuracy versus inference time per window (averaged over validation and test) across four subjects. Marker shape denotes subject, color denotes model; outlined markers are Pareto-optimal for their subject. Parenthetical numbers in the legend count the subjects for which each model is Pareto-optimal. The traditional ASR baseline (m01) is Pareto-optimal for none (0/4) and requires $>20\times$ more inference time.}
    \label{Fig5}
    \vspace{-10pt} 
\end{figure}
\begin{figure*}[!t]
    \centering
    \includegraphics[width=0.90\textwidth]{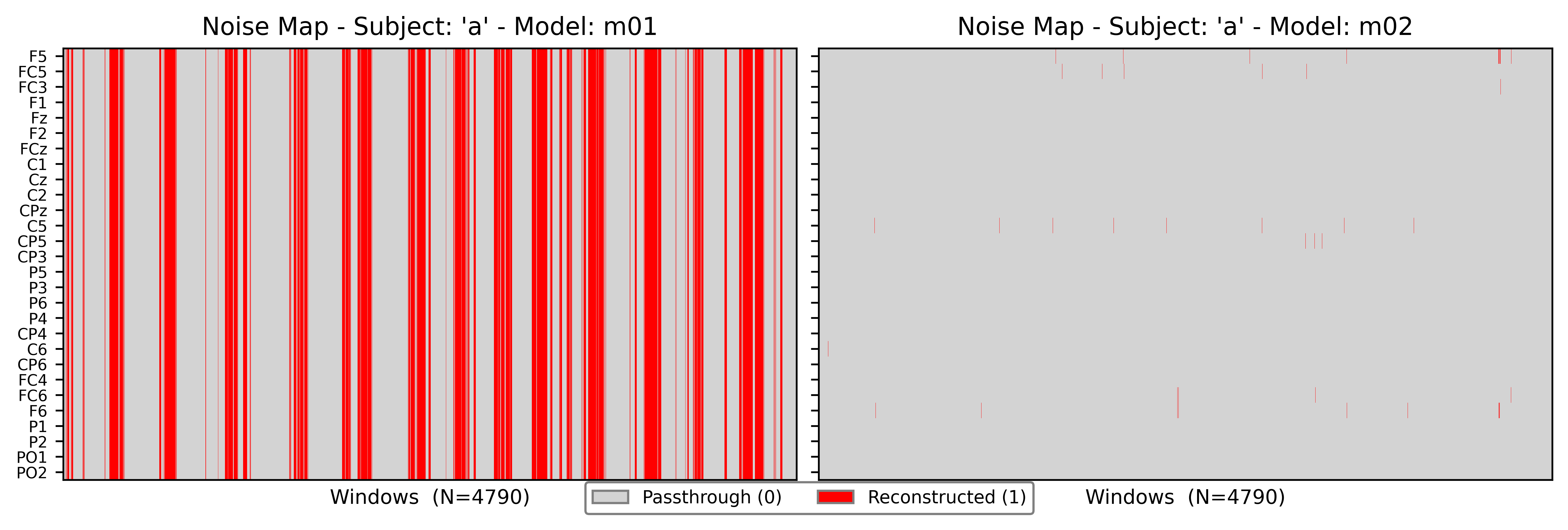}
    \caption{Channel-wise noise map ($N=4790$ noisy windows) for Subject~a.
    Each row represents an EEG channel, and each column represents a sliding
    window. Red indicates reconstructed channels; gray indicates passthrough
    channels. For m01, a channel is flagged when the absolute difference
    between the raw and ASR-reconstructed signals exceeds $10^{-5}$ at any
    time point. Left: traditional ASR (m01). Right: nASR (m02).}
    \label{Fig6}
    \vspace{-10pt} 
\end{figure*}
\subsection{Training Convergence} 
Fig.~\ref{Fig4} shows the best epoch per model, with bubble size encoding test accuracy. The nASR variants converged consistently within the Good zone (5--15 epochs) across all subjects, with m04 the slowest for `a' (epoch 14). The baseline m01 showed the longest convergence for `a' (epoch 27) and the highest test accuracy (0.640), yet this was comparable to nASR variants m03 (0.636) and m05 (0.628), which converged roughly 4$\times$ faster; notably, for `a', m05 showed the highest test F1-Score (0.573).

\subsection{Accuracy–Efficiency Trade-off: Pareto Analysis} 
Fig.~\ref{Fig5} evaluates the trade-off between balanced accuracy and inference time per window (both averaged over validation and test) using a Pareto frontier across four subjects. The total inference time for the baseline m01 includes both traditional ASR processing and network inference, yielding an overall mean of $\sim$7.6 ms (balanced accuracy 0.585). In contrast, all nASR variants (m02--m05) operate within $\sim$0.2--0.5 ms, representing a $>20\times$ reduction in inference time at comparable accuracy. Among the nASR variants, m03 (0.36 ms, balanced accuracy: 0.61) and m02 (0.34 ms, balanced accuracy: 0.60) are the most frequently Pareto-optimal (3/4 subjects each), followed by m04 and m05 (2/4 each). Overall, these results demonstrate that the learnable nASR formulation achieves the baseline's accuracy at a fraction of its inference cost, with m02 and m03 being the most frequently Pareto-optimal across subjects. 
\vspace{-5pt} 
\section{DISCUSSION}
The lowest accuracy of m01 in most cases suggests that traditional ASR, while suppressing artifacts during preprocessing, may overcorrect EEG signals and discard task-relevant neural information. In contrast, the learnable nASR variants demonstrated more balanced validation-to-test performance, indicating that end-to-end optimization of the artifact-rejection threshold allows the model to preserve discriminative signal components.
\\
As shown in Fig.~\ref{Fig6}, the noise-mask comparison for Subject `a' highlights the limitations of traditional ASR (threshold fixed at 20): m01 reconstructed over 30\% of noisy windows across all channels, leading to excessive signal modification and loss of task-relevant neural features. In contrast, the nASR variant m02 flags only a small subset of windows as noisy, and within those, only the primary artifact-contributing channels are reconstructed. This demonstrates that nASR identifies an effective threshold and isolates the main artifact sources at the channel level, thereby preserving task-relevant neural dynamics.
\\
Furthermore, the $>20\times$ reduction in inference time achieved by all nASR variants relative to m01 is a practically significant result, confirming nASR as a strong candidate for real-time BCI applications where both decoding accuracy and low latency are critical.
\\
The ablation results provide further insight into individual component contributions. Covariance estimation over the full overlapping window (m04) ranked among the weakest nASR variants, confirming that incorporating overlapping segments into covariance estimation is redundant and can be detrimental to performance. The neighborhood-based reconstruction (m03) and the global-mean reconstruction (m02) were the most consistently Pareto-optimal variants (3/4 subjects each), performing comparably overall. The marginal cases where m03 trailed m02 suggest that spatially proximate clean neighbors may still carry correlated residual noise, limiting the benefit of locality-constrained reconstruction over a broader channel average. This motivates our future work, which aims to replace simple channel averaging with a generative AI-based reconstruction strategy via EEGReXferNet~\cite{Sarkar2025}, to better recover artifact-contaminated channels while preserving task-relevant neural dynamics.
\\
Traditional ASR was tested only at its default threshold (20); a comparison across varied ASR settings is needed to fully characterize nASR's advantage. In addition, the present evaluation is limited to four subjects from the BCI Competition IV (Dataset 1). While the results are consistent across subjects, broader validation across a larger cohort, multiple datasets, and diverse BCI paradigms is needed to confirm the generalizability of the nASR framework.
\vspace{-5pt} 
\section{CONCLUSIONS}
This paper presented nASR, a novel trainable neural layer for channel-level EEG artifact subspace reconstruction with learnable dual thresholds. Ablation results confirm that nASR outperforms traditional ASR on classification accuracy while achieving $>20\times$ reduction in inference time, making it a strong candidate for real-time BCI.
\vspace{-1pt} 
\section*{Code Availability}
Source code available at: \href{https://github.com/ShanSarkar75/nASR}{github.com/ShanSarkar75/nASR}.

\section*{Acknowledgment}
The authors thank the BCI Competition IV organizers for 
providing the publicly available dataset used in this study.

\end{document}